%% file: capitanio2011.tex
\documentclass[useAMS,usenatbib]{mn2e}
\usepackage{graphicx}
\usepackage{subfigure}
\usepackage{longtable}\usepackage{lscape}
\usepackage{times}
\usepackage{epsfig}
\DeclareGraphicsExtensions{.pdf, .jpg}
\title[The peculiar IGR~J17091$-$3624 2011 outburst]{The peculiar 2011 outburst of the black hole candidate IGR~J17091$-$3624, a GRS~1915$+$105 like source?}

\author[F. Capitanio et al.]{F. Capitanio$^{1}$\thanks{E-mail:
fiamma.capitanio@iasf-roma.inaf.it}, M. Del Santo$^{1}$, E. Bozzo$^{2}$, C. Ferrigno$^{2}$, G. De Cesare$^{1}$, A. Paizis$^{3}$\\
$^{1}$Istituto Nazionale di Astrofisica, IAPS, Via Fosso del Cavaliere 100, 00133 Rome, Italy\\
$^{2}$ISDC Data Centre for Astrophysics, Chemin d'Ecogia 16, 1290 Versoix, Switzerland\\
$^{3}$Istituto Nazionale di Astrofisica, IASF-Mi, Via Bassini 15, I-20133 Milano, Italy} 

\begin{document}

\date{Accepted 2012 February 28. Received 2012 February 24; in original form 2011 October 19}

\pagerange{\pageref{firstpage}--\pageref{lastpage}} \pubyear{2011}

\maketitle

\label{firstpage}

\begin{abstract}
We report  on the long-term monitoring campaign of the black hole candidate IGR~J17091$-$3624 performed with {\it INTEGRAL} and {\it Swift} during the peculiar outburst started on January 2011. 
We have studied the  two month spectral evolution of the source in detail.  Unlike the previous outbursts, the 
initial transition from the hard to the soft state in 2011 was not followed by the standard spectral evolution expected for a transient black hole binary. 
IGR~J17091$-$3624 showed pseudo periodic flare-like events in the light curve, closely resembling those observed from  GRS~1915$+$105. 
 We find evidence that these phenomena are due to the same physical instability process ascribed  to GRS~1915$+$105. 
Finally we  speculate that the  faintness of IGR~J17091$-$3624 could be not only due to the high distance of the source but to the high inclination angle of the system as well.

\end{abstract}


\section{Introduction}
\label{intro}
The Black Hole Candidate (BHC) IGR~J17091$-$3624 was discovered   by {\it INTEGRAL}/IBIS during a Galactic Centre observation on 2003 April 14--15~\citep{Atel1}. 
At the onset of the   discovery outburst, the source showed a  hard spectrum with a flux of about $\sim$20 mCrab in the 40--100 keV energy range.  
The analysis of IBIS, JEM-X, and {\it RXTE}/PCA data of the
whole outburst   ~\citep{Cap,Lut,Lut2} revealed an indication of a hysteresis-like  behaviour.  The presence of
a hot disc blackbody emission component during the softening of the X-ray emission of the source  was also unveiled. 

After the {\it INTEGRAL} discovery,   IGR~J17091$-$3624 was searched in the archival data of both    TTM-KVANT~\citep{Atel2} and    {\it BeppoSAX}/WFC~\citep{Atel4}. In the former archive, one outburst was discovered dating back to 1994 and 
reaching a flux of 10 mCrab in the 3-30 keV energy band; the analysis of   {\it BeppoSAX}/WFC data revealed that a second outburst had occurred in 2001,    reaching a flux of 
14$\div$20 mCrab (2-10 keV).

IGR~J17091$-$3624 lies at 9.6$\arcmin$ from another transient X-ray binary, IGR~J17098-3628,   discovered on 2005 March 24 ~\citep{Atel444} when it underwent a 4 year  long outburst~\citep{Cap2}. 
On 2006 August 29 and 2007 February 19, two {\it XMM-Newton} observations of the region around these two sources were performed. While IGR~J17098-3624 was detected    in a relatively bright state in both observations,  IGR~J17091$-$3624     was not detected and an X-ray upper limit of  7$\times$10$^{32}$ erg s$^{-1}$ was obtained \citep[assuming a distance of 8 kpc;][]{Cap2}. 

  The refined position of IGR~J17091$-$3624 provided by 
\citet{ATEL1140} ruled out the tentative radio counterpart previously proposed for the source~\citep{Atel3,Pan}. A re-analysis of the  archival radio observations  performed  9 days after the source discovery by IBIS in 2003,  enabled the identification of a faint transient radio source  (sub-mJy level at 5 GHz)    that showed  a flux  increase    in  the subsequent two weeks and an  inverted spectrum,   a signature of a compact jet  ~\citep{Cap2}. This was consistent with the Low/Hard spectral state ( hereafter LHS)    observed by {\it INTEGRAL} in the same period~\citep{Cap}. The source behaviour during the 2007 observation campaign was typical of a BHC in outburst, even if the relatively low X-ray  flux of the source 
( 0.5--10\,keV peak flux of  $\sim$2 $\times$10$^{-9}$ erg~cm$^{-2}$~s$^{-1}$)  hindered a detailed spectral evolution study~\citep{Cap2}.

At the end of January 2011 the {\it Swift}/BAT hard X-ray transient monitor reported   a renewed activity from IGR~J17091$-$3624. The source flux increased   from 20 mCrab on January 28 up to 60 mCrab on February 3  in the energy range 15-50 keV~\citep{Atel3144,Atel3148}. The corresponding XRT spectrum  obtained with a ToO observation was well described by  
an absorbed power law with a photon index of  1.73$\pm$0.29 ~\citep{Atel3148}.
On 2011 February 7, the region around IGR~J17091$-$3624 was also observed by the IBIS/ISGRI and JEM-X telescopes on board the {\it INTEGRAL} satellite. The estimated source flux in the 20-100 keV energy range was 120 mCrab. The combined ISGRI+JEM-X spectrum (5-200 keV) could be well described by an absorbed cut-off power law model with a photon index of $\sim$1.4 and a  high energy cutoff of about 110 keV. This suggested that  the source was in LHS ~\citep{Atel3159}.

Follow-up radio observations carried out with the ATCA telescope   measured   a flat spectrum \citep{Atel3150,Atel3167,Rodriguez} associated 
with  self absorbed compact jets, as expected in accreting black holes in the LHS. Later on, \citet{Rodriguez} reported also on 
the detection of a discrete jet ejection event  usually observed when a BHC undergoes a transition from 
the Hard Intermediate State ( hereafter HIMS) to the Soft Intermediate State ( hereafter SIMS). 
A 0.1 Hz QPO, increasing in frequency with the source flux and spectral softening, was revealed  by both ~\citet{Atel3168} and 
\citet{Atel3179}. These findings motivated a long monitoring campaign that was carried out with {\it Swift}/XRT, starting on February 28. 
The XRT observations were planned to be simultaneous with the {\it INTEGRAL} pointings  already scheduled in the direction of the source, 
in order to ensure the broadest possible energy coverage (0.3-200 keV) during the entire outburst.

As reported by~\citet{Atel3203}, on February 28 the XRT+IBIS joint spectrum resulted in a typical   High Soft State (HSS) shape, 
with a prominent disc black body component (kT$_{in}\sim$1keV)  and a power-law photon index of 2.2$\pm$0.2. No high-energy cut-off was present up to 200 keV.
On 2011 March 14  (MJD 55634)  a $\sim$ 10 mHz QPO was  detected in a 3.5 ks {\it RXTE} observation~\citep{Atel3225}. One week later, {\it RXTE}/PCA showed a continuous progression of quasi-periodic flare-like events occurring at a rate between 25 and 30 mHz. This kind of variability resembles the ``heartbeat'' variation observed in the Black Hole (BH) binary GRS~1915$+$105 \citep{Atel3230,Atel3418,Pahari}. 
 \citet{Altamirano} reported a detailed study of the behaviour of the flare-like events of IGR~J17091$-$3624 during the first 180 days of the outburst. This study  classified the different types of flares with the same scheme used by~\citet{Belloni_a} for GRS~1915$+$105.

In this paper we report on the {\it Swift} and {\it INTEGRAL} data analysis of the new outburst of IGR~J17091$-$3624 started at the end of  January 2011.

\section{Data reduction and analysis}
\label{data}

The XRT ToO follow-up observations were performed, when possible, simultaneously to the {\it INTEGRAL} ones~\citep{Atel3159}. {\it INTEGRAL} data were collected in the framework of the  Galactic bulge observations\footnotemark\footnotetext[1]{http://integral.esac.esa.int/BULGE} (public data) and the open time observation  of the  RX J1713.7-3946 field. 
Due to the long duration of the outburst, {\it Swift}/XRT data were collected also in the period in which the region around IGR~J17091$-$3624 became unobservable by {\it INTEGRAL}. 
In this paper we made use of the whole available  data set of   {\it INTEGRAL} and {\it Swift} observations performed from 28 January to 14 August  2011.

 The XRT observations were taken in window timing mode in order to avoid the pile-up  effects. Each observation was composed of two or more segments. 
We reported only the analysis of  the first segments of all XRT observations,  since the other segments were always consistent with the first  segments of each observation. 
For the XRT data analysis we followed standard procedures~\citep{Burrows} and the technique summarized in ~\citet{Bozzo}.
 XRT light curves and the hardness-intensity diagrams were obtained  from the XRT data  extracting two different energy ranges, 0.3-4 keV and 4-10 keV. 
 
For the {\it INTEGRAL} data analysis, we used the latest release
of the standard Offline Scientific Analysis, OSA version 9.0,  distributed by the ISDC~\citep{Courvoisier} and the latest response matrices available. 
In particular, the IBIS response matrices were produced using the closest available Crab observations to the 2011 outburst of IGR~J17091$-$3624.
Our {\it INTEGRAL} analysis was focused on ISGRI~\citep{Lebrun}, the low-energy detector of the $\gamma$-ray 
telescope IBIS ~\citep{Ubertini} and on the X-ray monitor JEM-X~\citep{Lund}.
 Unfortunately, due to the {\it INTEGRAL} observing strategy combined to the small JEM-X field of view (FOV), 
  IGR~J17091$-$3624 was not in the JEM-X  {\it FOV}  in most of the observations.
During the {\it INTEGRAL} observations both JEM-X modules were switched on.  However, 
for the data analysis we used the second module (JEM-X2) and checked the consistency with module~1. 
The ISGRI and JEM-X spectra were extracted in    20-200 keV  and  3-20 keV, respectively. 
A systematic error of 2\% was taken into account for spectral analysis ~\citep[see also][]{Jourdain}. 

Details on all the {\it Swift} and {\it INTEGRAL} data analysed in this paper are  given in  Table~\ref{parameters} (columns 1-4).
 The spectral and timing analysis have been performed with HEASOFT 6.9 package. In particular, the periods of the flare-like events were calculated with the FTOOL {\it efsearch}. 
The {\it rms} values were estimated  from the source light-curves by using an {\it ad hoc} 
developed tool and {\it the IDL Astronomy User's Library procedures}\footnotemark\footnotetext[2]{{http://idlastro.gsfc.nasa.gov/}}.
   For the {\it rms} calculation,  we divided the light curves, extracted in 1\,s bins,  into 140\,s chunks. For each segment we computed the fractional {\it rms}  after subtracting the expected white noise. We then estimated the fractional {\it rms} of the light curves and its uncertainty from the average and standard deviation of the single determinations. The effective frequency range over which the {\it rms} is integrated is therefore 0.007-0.5.

\scriptsize
\onecolumn
\begin{landscape}
\begin{longtable}{cccccccccccc}
\caption[Observations log and spectral parameters of the outburst evolution]{ Observations log and spectral parameters of the outburst evolution.
Note: all the errors are at 90\% confidence level.   N is the label of each XRT observation associated to the points of Figure~\ref{HID} and Figure~\ref{RMS}; ID is the XRT observation number; Date is the date of the XRT observation; {\it rms} is the value of the root-mean-square amplitude of each XRT observations averaged in an interval between 0.007 and 0.5 Hz.  {\it INTEGRAL} REV indicates, when available, the revolution number of {\it INTEGRAL} simultaneous observations; T$_{in}$ is the inner temperature of the {\it diskbb} model in {\tt XSPEC}; NORM {\it diskbb} is the normalization of the {\it diskbb} model  proportional to the square of the inner disc radius square; $\Gamma$ is the  power law photon index and E$_{c}$ is the high energy cut off; FLUX$_{(2-10)keV}$ is the unabsorbed flux between 2 and 10 keV.}\label{parameters} \\
\multicolumn{1}{c}{{N}}
&\multicolumn{1}{c}{{ID}} 
& \multicolumn{1}{c}{{Date}} 
& \multicolumn{1}{c}{{XRT EXP}} 
&\multicolumn{1}{c}{{{\it INTEGRAL}}}
&\multicolumn{1}{c}{{{\it rms}}}
&\multicolumn{1}{c}{{T$_{in}$}}
&\multicolumn{1}{c}{{NORM}}
&\multicolumn{1}{c}{{$\Gamma$}}
&\multicolumn{1}{c}{{E$_{c}$}}
&\multicolumn{1}{c}{{FLUX$_{(2-10)keV}$}} &\multicolumn{1}{c}{{$\chi^{2}_{red}$(d.o.f.)}}\\

\multicolumn{1}{c}{{-}}
&\multicolumn{1}{c}{{-}} 
& \multicolumn{1}{c}{{MJD}} 
& \multicolumn{1}{c}{{s}} 
&\multicolumn{1}{c}{{REV}}
&\multicolumn{1}{c}{{cnt}}
&\multicolumn{1}{c}{{keV}}
&\multicolumn{1}{c}{{\it diskbb}}
&\multicolumn{1}{c}{{-}}
&\multicolumn{1}{c}{{keV}}
&\multicolumn{1}{c}{{ ($\times$10$^{-10}$erg~cm$^{-2}$s$^{-1}$)}}
&\multicolumn{1}{c}{{-}}\\ 
 \hline
\hline
\endfirsthead

\multicolumn{3}{c}
            {\footnotesize\itshape\tablename~\thetable:
continued from previous page} \\

\multicolumn{1}{c}{{N}}
&\multicolumn{1}{c}{{ID}} 
&\multicolumn{1}{c}{{Date}} 
&\multicolumn{1}{c}{{XRT EXP}}
&\multicolumn{1}{c}{{{\it INTEGRAL}}}
&\multicolumn{1}{c}{{{\it rms}}}
&\multicolumn{1}{c}{{T$_{in}$}}
&\multicolumn{1}{c}{{NORM}}
&\multicolumn{1}{c}{{$\Gamma$}}
&\multicolumn{1}{c}{{E$_{c}$}}
&\multicolumn{1}{c}{{FLUX$_{(2-10)keV}$}}
&\multicolumn{1}{c}{{$\chi^{2}_{red.}$(d.o.f.)}}\\ 

 \multicolumn{1}{c}{{-}}
& \multicolumn{1}{c}{{-}}
& \multicolumn{1}{c}{{MJD}} 
& \multicolumn{1}{c}{{s}} 
&\multicolumn{1}{c}{{REV}}
&\multicolumn{1}{c}{{cnt}}
&\multicolumn{1}{c}{{keV}}
&\multicolumn{1}{c}{{\it diskbb}}
&\multicolumn{1}{c}{{-}}
&\multicolumn{1}{c}{{keV}}
&\multicolumn{1}{c}{{($\times$10$^{-10}$erg~cm$^{-2}$s$^{-1}$)}}
&\multicolumn{1}{c}{{-}}\\ 
 \hline
\hline
\endhead
 \multicolumn{3}{c}{{Continued on next page}} \\ 
\endfoot
\endlastfoot
\input{tabella_referee.tex}
\hline
\hline
\end{longtable}
\end{landscape}
\twocolumn
\normalsize

\section{Results}
\label{evo}
  The 2011 outburst of IGR J17091-3624 can be divided in two main phases: during the first one,  the source underwent the typical sequence of events of a transient BH in outburst (described in Section~\ref{evo_1}); during the second part, it exhibited ``heartbeat'' variability previously observed only in GRS~1915$+$105 (Sections~\ref{heartbeat_p} and ~\ref{heartbeat_p2}). Finally, a detailed study on the presence of a Compton reflection component and iron line upper limit are given in Section~\ref{reflection}.

 \subsection{The initial phases of the outburst}
\label{evo_1}
\begin{figure}
  \includegraphics[angle=0, scale=0.3] 
 {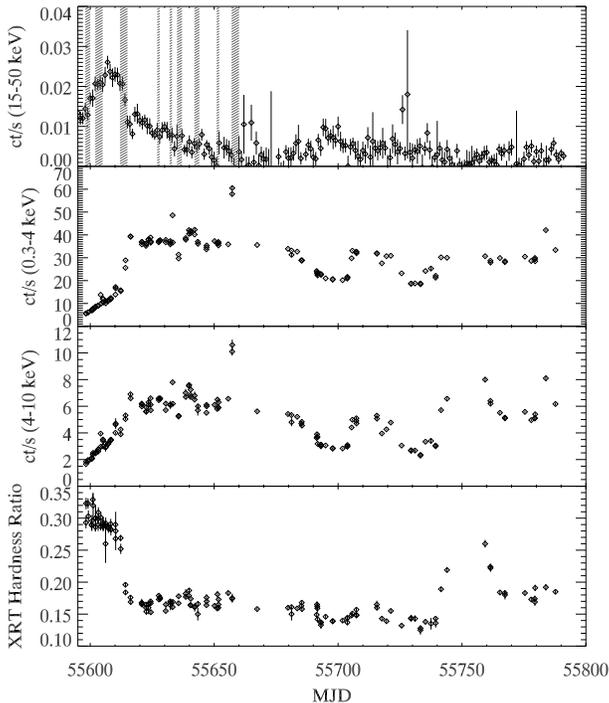}
\hspace{5cm}
\caption{Top panel:  {\it Swift}/BAT (15-50 keV) count rate (bin time= 1 day). The shadowed parts represent the {\it INTEGRAL} observation periods. 
Second panel: XRT (0.3-4 keV) count rate (bin time =4000s). Third panel: XRT (4-10 keV) count rate (bin time =4000s). 
Bottom panel: XRT Hardness Ratio  (defined as the ratio between the 4-10 keV to the 0.3-4 keV count rate).}\label{xrtlctot}
\end{figure}

  The outburst  of IGR~J17091$-$3624  started on MJD$\sim$55598 (Figure~\ref{xrtlctot}) and in about 12 days the X-ray flux of the source (2-10 keV) increased of about 70\%.
During this starting phase, the {\it Swift} and {\it INTEGRAL} simultaneous data, when available, could be well fit by an absorbed cutoff power-law model. The source showed a typical hard state spectrum and the  photon index and high-energy cutoff remained consistent within the errors ($\Gamma\sim$1.5, E$_{c}\sim$100 keV, see Table\ref{parameters} for details). 
The equivalent hydrogen column density value was consistent with the one reported by~\citet[][]{Atel3148}, N$_{H}$=(1.1$\pm$0.3)$\times$10$^{22}$cm$^{-2}$. 

\begin{figure}
  \includegraphics[angle=-90, scale=0.3] {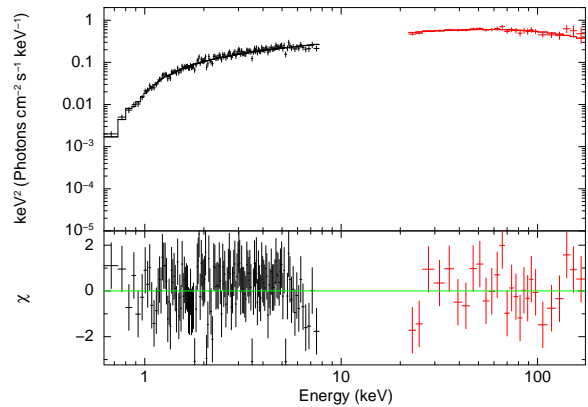}
     \caption{{\it Swift}/XRT and {\it INTEGRAL}/IBIS joint unfolded spectrum at the beginning of the outburst.  The source presents a typical  LHS spectrum (observation n$^{o}$3 in Table~\ref{parameters}).}
         \label{LHS}
   \end{figure}

 Figure~\ref{LHS} shows the  combined XRT-ISGRI unfolded LHS spectrum  along with the residuals expressed 
in terms of sigmas (MJD=55603.2, observation n$^{o}$6 in  Table~\ref{parameters}).

On MJD$\sim$55610.2, the source displayed evidence for   a beginning of a spectral transition to the softer state. The flux continued to increase more rapidly: $\sim$100\% from the observation n$^{o}$12 until the observation n$^{o}$15 (about 6 days). But, this time, a significant softening of the hard X-ray spectrum (see i. e. bottom panel of Figure~\ref{xrtlctot}) was observed, together with a drop in the hard X-ray flux. 

\begin{figure}
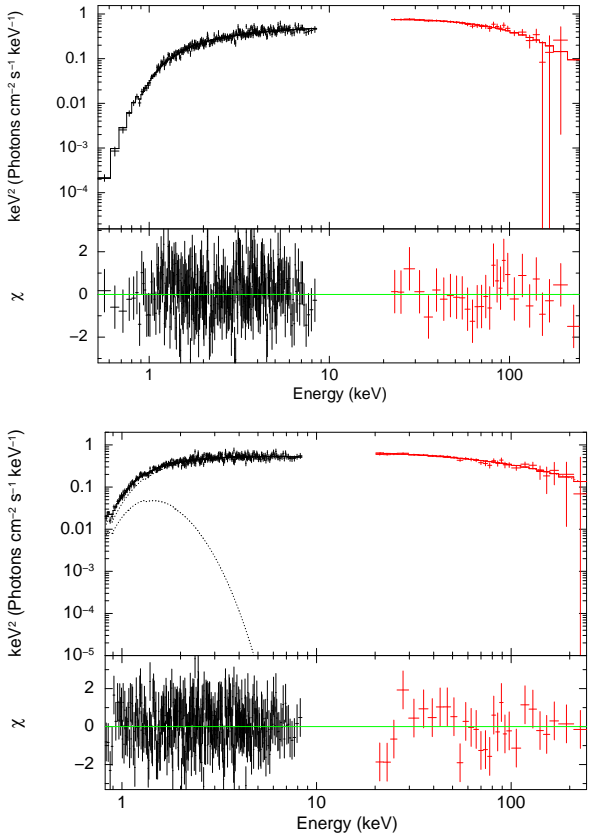

\includegraphics[angle=-90,scale=0.3]{capitanio11_fig3.ps}
\includegraphics[angle=-90, scale=0.3] {capitanio11_fig4.ps}
     \caption{Two {\it Swift}/XRT and {\it INTEGRAL}/IBIS joint intermediate spectra during the transition from  the Low Hard State (LHS) to the High Soft State (HSS). The two spectra have been collected from data separated by 2 days. Top spectrum: observation n$^{o}$13 in Table~\ref{parameters}. Bottom spectrum: observation n$^{o}$14 in Table~\ref{parameters}.}
         \label{HIS}
   \end{figure}

During the transition the spectra became steeper and in about two days the fit required a multicolor disc blackbody 
component  \citep[  modeled with {\tt diskbb} in {\tt XSPEC},][ hereafter {\tt MDBB}]{Mitsuda}.
Figure~\ref{HIS} shows two spectra extracted at the intermediate hardness values (HR$\sim$0.2, observation n$^{o}$13 and n$^{o}$14). 
  An acceptable fit to these spectra could be obtained by using an absorbed cutoff power law model).
Adding the {\tt MDBB} component, the F-test probability of a chance improvement is $~$7\% and  $~$0.4\%, for the observations n$^{o}$13 and n$^{o}$14, respectively.
  Thus it is reasonable to add a {\tt MDBB} component only to the second spectrum.\\
The obtained spectral parameters of the spectrum n$^{o}$14 are compatible with  the intermediate spectral states of a BHC  (see e. g.~\citet{Fender} and \citet{Remillard} and references therein).
During   this transition from the hard to the soft state, the inner temperature of the {\tt MDBB} component (kT$_{in}$) increased from 
0.3 keV (observation n$^{o}$14) to $\sim$1 keV (observations n$^{o}$15$\div$16), 
while its normalization   decreased significantly\footnotemark\footnotetext[3]{ In the  {\tt MDBB} model ~\citep{Mitsuda} the square root of the 
normalization constant is proportional to the apparent inner radius of the truncated disc. However, when the high energy 
behaviour of the spectrum is modeled with a power law component, the evolution of the disc internal radius can be significantly underestimated~\citep[see e. g.][p. 28-29]{Done}.}.

\begin{figure}
  \includegraphics[angle=-90, scale=0.3] {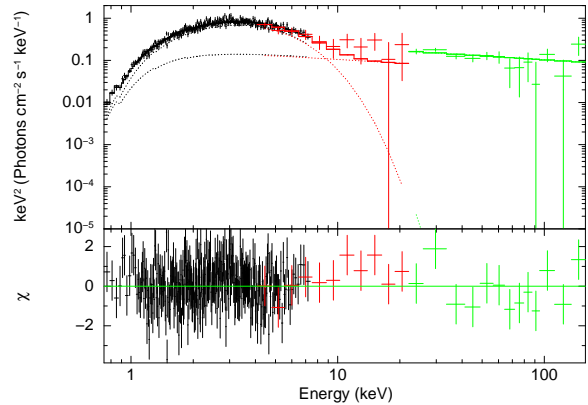}
     \caption{ {\it Swift}/XRT {\it INTEGRAL}/JEM-X2 and {\it INTEGRAL}/IBIS unfolded  spectra of the IGR~J17091$-$3624 soft state (see Section~\ref{disc}). 
      The fit is an absorbed {\tt MDBB} plus a power law. No reflection component is needed in the fit (the spectral parameters values are reported in Table ~\ref{parameters}, 
      observation n$^{o}$33).}
         \label{HIS2}
   \end{figure}

 At the end of the transition to the soft state   (observation n$^{o}$16), the disc temperature reached a value of about 1 keV, while the power law photon index  reached $\sim$2.1, with no cutoff detectable up to about 200 keV (see Table ~\ref{parameters} for details).  The fractional {\it rms} amplitude of the X-ray emission from IGR~J17091$-$3624 as measured by XRT data decreased from previous values (25$\div$30\%) up to about 4$\div$5\% (see Figure~\ref{RMS}). Thus, as also reported by~\citet{Atel3203}, the source is probably in the HSS.
 In the following 65 days (until observation n$^{o}$42) 
the spectral characteristics of the source showed no significant variability.
Figure~\ref{HIS2} shows the unfolded spectrum of IGR~J17091$-$3624 after the transition (observation n$^{o}$33).  
The fit to these data was obtained with an absorbed {\tt MDBB} plus a simple power law component. 
  No Compton-reflection from the disc surface and no iron line models were  required by the data even though 
 these components are usually expected to be very strong in the canonical soft state of BH binaries~\citep{Gierlinsky}.  

On MJD=55655.8 (observation n$^{o}$34) a short flare, reaching a peak flux of 3$\times$10$^{-9}$~erg~cm$^{-2}$~s$^{-1}$ (2-10 keV) was detected.  
No significant changes in the spectral properties of the source were detected during this event.

\subsection{The appearance of the ``heartbeat''}
\label{heartbeat_p}
\begin{figure*} 
 \centering
\includegraphics[angle=0,scale=0.5]{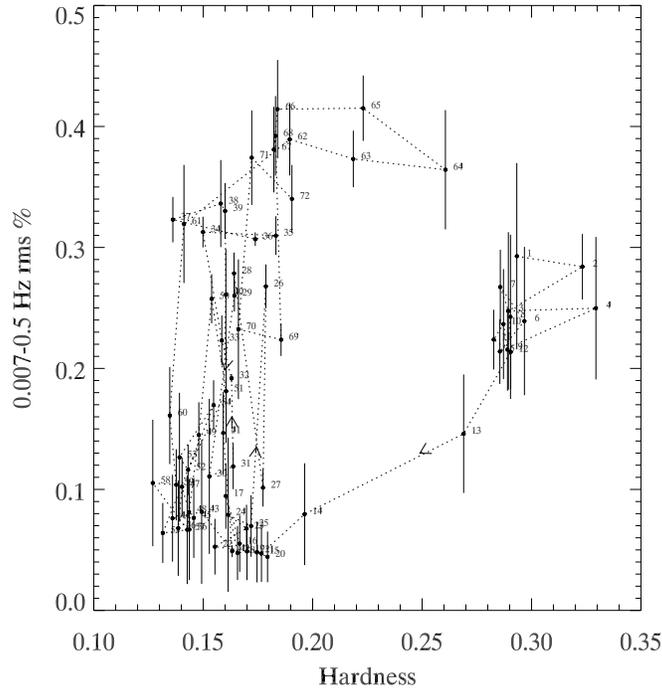}
     \caption{Hardness-rms diagram of each XRT pointing of the IGR~J17091$-$3624 outburst. For the observations with more than one segment only the first  one has been considered. For the usage of {\it rms} as a tracer of the different accretion regimes see e. g.~\citet{Munoz-D} and~\citet{Cap3}. In order to get a more readable Figure, we did not show the hardness error bars that are, instead, reported in Figure~\ref{HID}.}
         \label{RMS}
\end{figure*}

 Figure~\ref{RMS} shows the fractional {\it rms} amplitude as a function of the hardness ratio, hereafter HR\footnotemark\footnotetext[4]{ We defined as the hardness ratio the ratio of the counts in the  4-10 keV energy band to the counts in the 0.3-4 keV energy band  in each XRT observation.}.  As mentioned above, during the transition from the hard to the soft state, the fractional {\it rms} and the   HR decreased as expected by a typical transient BH entering  the HSS~\citep{Fender}.

\begin{figure}
     \centering
\includegraphics[angle=-90,scale=0.28]{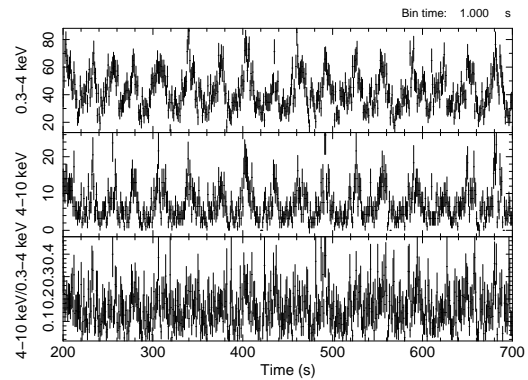}
     \caption{Zoom of the XRT count rate of the  observation  n$^{o}$28 in Table~\ref{parameters}.   The time bin is 1\,s and the start time is  MJD=55640.5.}
         \label{hbeats}
   \end{figure}

\begin{figure}
\includegraphics[angle=0,scale=0.2]{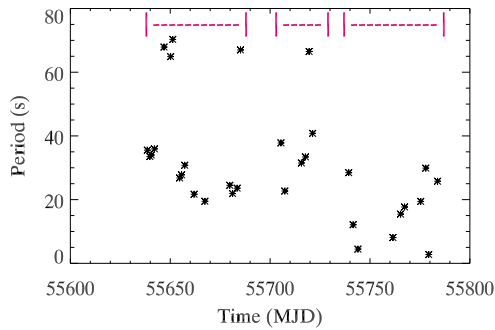}
     \caption{ ``heartbeat'' period versus time. The dashed segments represent the three different groups of observations discussed in the text.}
         \label{periods}
\end{figure}

\begin{figure}
\begin{center}
\includegraphics[angle=0,scale=0.2]{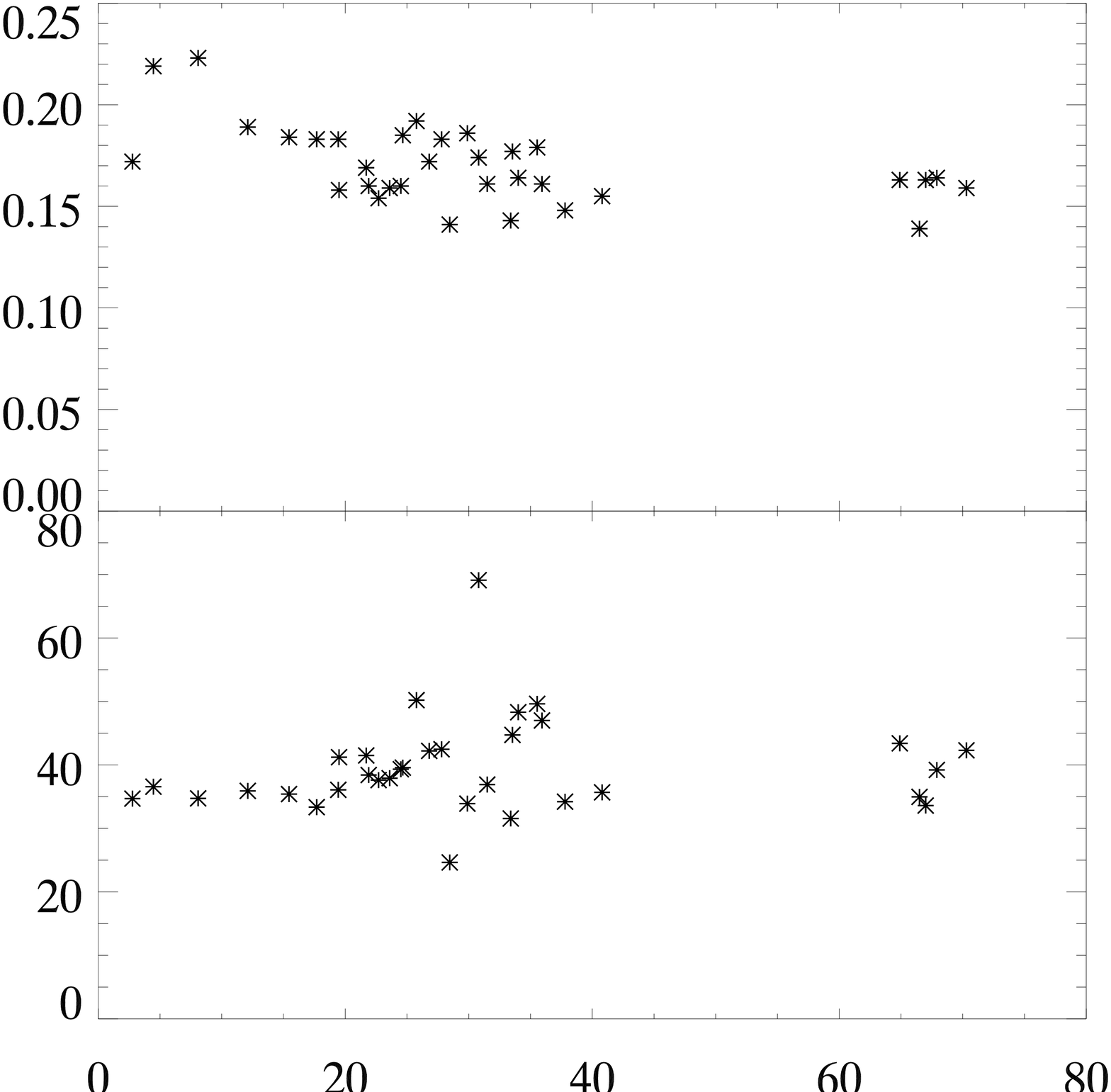}

\hspace{5cm}
     \caption{Top panel:   hardness versus ``heartbeat'' period. Bottom panel: XRT count rate versus ``heartbeat'' period. }
         \label{periods2}
\end{center}
\end{figure}

However,  from observation n$^{o}$ 26
the fractional rms amplitude moved away from the expected values and started  to increase and decrease rapidly with a chaotic behaviour (see e. g. Figure~\ref{RMS}). The rapid increases correspond to the observations in which the quasi-periodic flare-like events are detected in the light curves (the ``heartbeat'' in analogy with GRS~1915$+$105, see also Section~\ref{intro}).  
As an example, Figure~\ref{hbeats} shows a zoom of the light curve of one of the XRT observations in which the ``heartbeat'' is detected. 
 The ``heartbeat'' oscillations vary in intensity and in hardness; in some observations they are not detected at all (in these cases lower values of the fractional rms amplitude are measured).   No significative variations can be observed in the spectra of each XRT observation with or without the presence of the ``heartbeat''. We also observed that the flare-like events lose coherence and change  their period with time. Figure~\ref{periods} shows the evolution of the ``heartbeat'' period as a function of   time. 
This behaviour is consistent with what observed with {\it RXTE}~\citep{Atel3299,Altamirano}.
The two panels of Figure~\ref{periods2} show the ``heartbeat'' period as a function of hardness and XRT count rate, respectively.  No evident correlation between the periods of the flare-like events with the count rate or the HR has been found. The only peculiarity is the presence of a sort of "forbidden zone" in the possible period values (from $\sim$40\,s to $\sim$65\,s, Figures~\ref{periods},~\ref{periods2}). For a detailed discussion of the different ``heartbeat'' states of IGR~J17091$-$3624 see~\citet{Altamirano}.

  No significant  detection of the ``heartbeat'' was found in the IBIS light curve because of the faintness of the source in the hard X-ray domain (20--200\,keV) 
and  the relatively poor statistics.  

After   MJD$\sim$55690 (observations n$^{o}$ 43-44), the ``heartbeat''  was no longer detected  
and at the same time the flux in the 15--50\,keV energy band started to increase again (see the BAT light curve in Figure~\ref{xrtlctot}).  The spectral 
analysis of the observations collected during this period showed that the inner temperature of the {\tt MDBB} component decreased  down to $\sim$1~keV
and a power-law  component was also required in order to have an acceptable fit  of the XRT spectra. In the previous observations, a power law component additional to {\tt MDBB} was required only when XRT and IBIS data were fitted simultaneously.
 Between observations n$^{o}$37-41 the {\it INTEGRAL} data were unavailable, 
and thus we could not constrain the properties of the source emission in the hard X-ray domain.

 On MJD=55705.6 (observation n$^{o}$49), the 15--50\,keV light curve started to decrease again. Correspondingly, 
the soft XRT light curve increase significantly (see Figure~\ref{xrtlctot}) and the XRT spectra reached again
approximately the same shape observed during the previously detected soft state.
On the same date, MJD=55705.6,  a second group of recurrent flare-like events appeared again in the light curves. 
At this time the flux variation of the flare events was less  pronounced and less coherent, while the periods scanned 
approximately the same range than in the previous group of events (see Figure~\ref{periods}).

 As Figure~\ref{xrtlctot} shows, from MJD$\sim$55730 until 55770 there was an increase of the XRT flux together with a sharp hardening.  
  The consequence of the hardening in the XRT spectra are an increase of the inner disc temperature and a decrease of the normalization constant of the {\tt MDBB} model, 
 NORM, that reached values of about $\sim$18 (see Table~\ref{parameters}). In particular NORM is proportional  to the square of the apparent 
 inner disc radius and to {\tt cos{\tt i}}, where {\tt i} is the angle between the disc and the observer~\citep{Mitsuda}.\footnotemark\footnotetext[5]{The connection 
 between the apparent inner disc radius and the inner radius itself  is reported by~\citet{Kubota}).} 
 Thus in order to obtain an inner radius with a plausible length, cos{\tt i} should be very small.   Simultaneously to the spectral hardening, 
the XRT light curves and  the corresponding power spectra clearly showed that a third group of recurrent 
flare-like  events started with a remarkably decreased period (see Figure~\ref{periods}).

 As an example, the four panels of  Figure~\ref{35-17flares} show  the XRT power spectra evolution, from MJD=55737.5 to MJD=55759.3  (observations n$^{o}$60$\div$64 ). This time interval corresponds to the reappearance of the flare-like events:
 at MJD=55737.3 (observation n$^{o}$60) there  were no flare-like events and the power spectrum presented   a power law like behaviour (panel {\it a}).
On MJD=55741.6 the flare-like events started again  and a prominent and broad feature appeared in the power spectrum shape (panels {\it b} and {\it c}). The frequency of this feature  changed with time  from  $\sim$0.72 Hz to  $\sim$0.22 Hz (panels {\it d} and {\it e}).

\begin{figure}
   \includegraphics[angle=-90, scale=0.3] {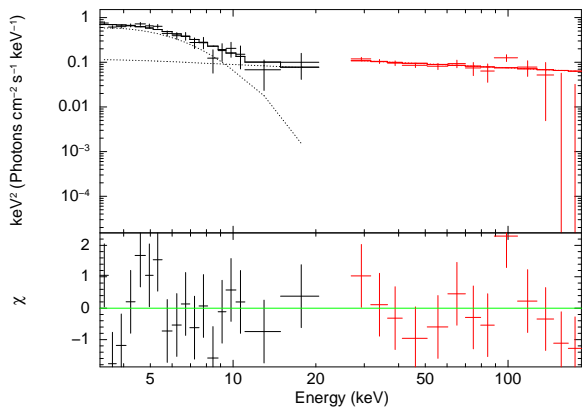}
 \caption{{\it INTEGRAL}/JEM-X2 and {\it INTEGRAL}/IBIS averaged spectrum of IGR~J17091$-$3624 
in the soft state during revolution 1078 (MJD=55785.0). The fit is an absorbed {\tt MDBB} plus a power law. The spectral parameters are reported in Table~\ref{parameters}. }
 \label{newspe}
  \end{figure}

The ``heartbeat'' is always detected during the final part of the {\it Swift} campaign with periods that span  from about 3\,s until 30\,s. The energy spectra of each single XRT observation were fitted with the same model than before  (absorbed multicolor disc black body plus a power law component). However,  after the observation n$^{o}$65, the inner temperature of the {\tt MDBB} component decreased from $\sim$1.5 keV to $\sim$1.3 keV (see Table~\ref{parameters} for details). The {\it INTEGRAL} observations, performed during revolution 1078 (MJD=55785.0), showed that the fit of the hard part of the spectrum is consistent with a simple power law component with a photon index of $\Gamma$=2.3$\pm0.2$ (see Figure~\ref{newspe}). 

\begin{figure*}
  \includegraphics[angle=0, scale=0.6] 
  {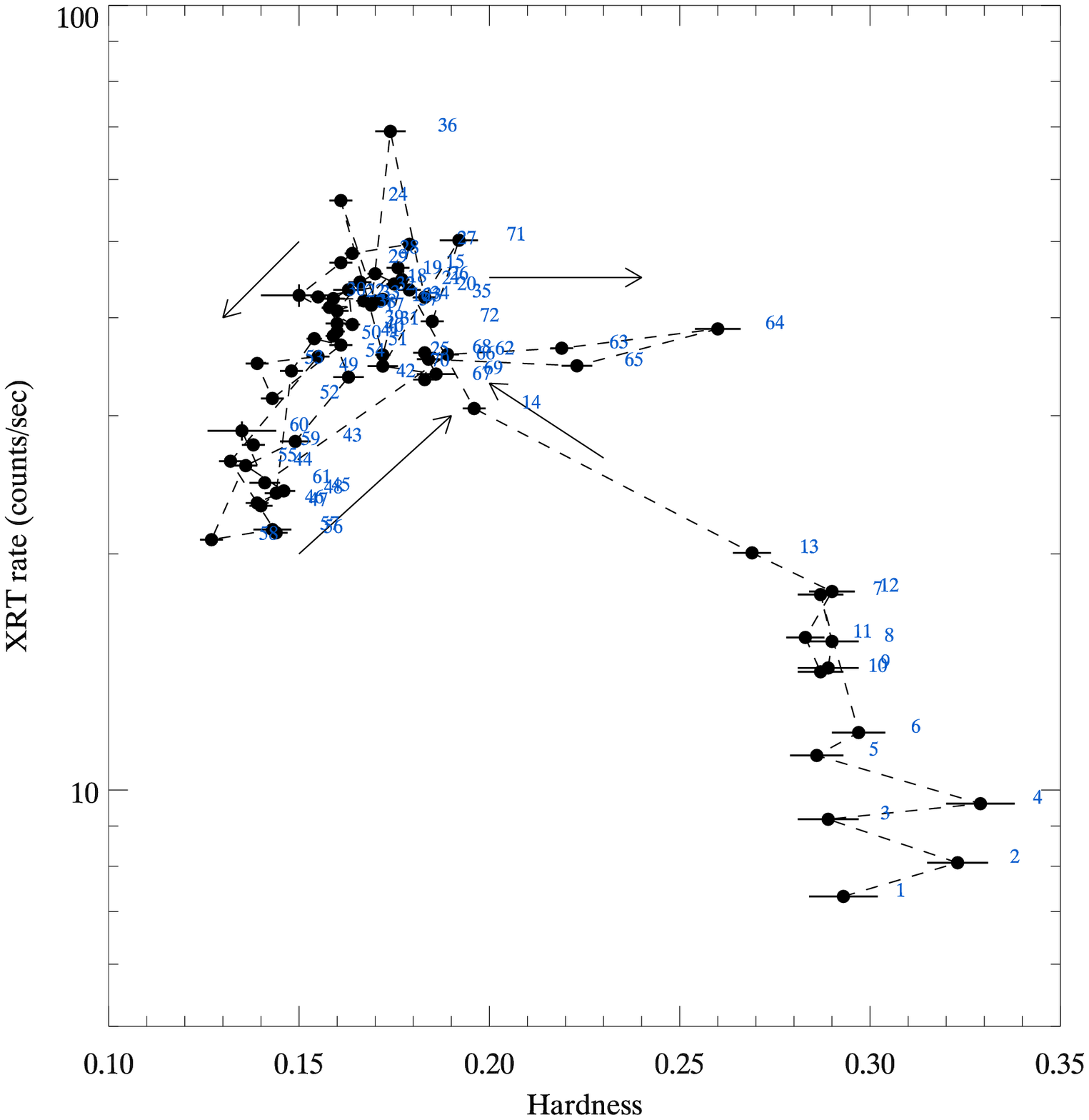}
\caption{ Hardness-intensity diagram (HID) of all the XRT 2011 outburst observations of IGR~J17091$-$3624.   
 For the observations with more than one segment only the first  one has been considered.
}
         \label{HID}
   \end{figure*}

 We also note that during the periods in which IGR~J17091$-$3624 displayed evidence for the ``heartbeat'' phenomena, its spectral evolution remained 
trapped in the top left corner of the hardness-intensity diagram (hereafter HID; see Figure~\ref{HID}) and outlined no more the canonical 
path through the different spectral states expected from a BHC in outburst (the so called q-track).

\begin{figure*}
     \centering
     \subfigure[id: 00035096016]
 {\includegraphics[angle=-90, scale=0.3] {capitanio11_fig10.ps}}
\subfigure[id: 00035096017]
{\includegraphics[angle=-90, scale=0.3] {capitanio11_fig11.ps}}\\
\subfigure[id: 00035096018]
{\includegraphics[angle=-90, scale=0.3] {capitanio11_fig12.ps}}
\subfigure[id: 00035096019]
{\includegraphics[angle=-90, scale=0.3] {capitanio11_fig13.ps}}\\
\subfigure[id: 00035096020]
{\includegraphics[angle=-90, scale=0.3] {capitanio11_fig14.ps}}
\caption{XRT power spectra evolution of  five observations (binned at 1s), from MJD=55737.5 to MJD=55759.3 (observation n$^{o}$60$\div$64), that correspond  to the reappearance of the flare-like events  of the last part of the XRT campaign of IGR~J17091$-$3624.}
         \label{35-17flares}
   \end{figure*}

\subsection{Spectra from the ``heartbeat'' }
\label{heartbeat_p2}
 In order to investigate the origin of the changes in the hardness ratio during the ``heartbeat'', we extracted XRT spectra in the time intervals corresponding to the highest ($>$60 ct/s) and lowest ($<$30 ct/s) count rates of the source spectra during the flaring activity.  For these data, we   performed a   rate resolved analysis adding-up time intervals corresponding to the  peaks and to the minima of the flare in each observation   (note, however, that the hardening of the different peaks was not constant; see for example the HR behaviour in Figure~\ref{hbeats}).   Because of the periodicity of the light curve, the rate resolved analysis overlaps with the phase resolved analysis.

A fit to the  spectra was obtained by using an absorbed {\tt MDBB} component. 
The spectral parameters at highest count rates  indicated a higher inner  disc temperature and  a hint for a smaller inner disc radius
(see Table~\ref{tab_peak} for details)  than what measured at   lower count rates.  
This behaviour is more  evident in   some observations of the first group of data showing recurrent flare like events (between MJD$\sim$55630 and MJD$\sim$55690),  where the flux variation during the flares was more pronounced. In fact, unfortunately, due to the low data statistics, only in a few observations was it possible to constrain the {\tt MDBB} normalization constant with enough confidence.

In the second group (from MJD$\sim$55700  to MJD$\sim$55730) the changes in the HR with the source count rates and the coherence of the ``heartbeat'' oscillation are less evident.  
We report in Table~\ref{tab_peak} the spectral parameters of three  
representative 
XRT observations selected at different time periods. 
The N$_{H}$ is fixed to be the same  for the different phases of the same observations.

\begin{figure}
 \includegraphics[angle=-90,scale=0.3]{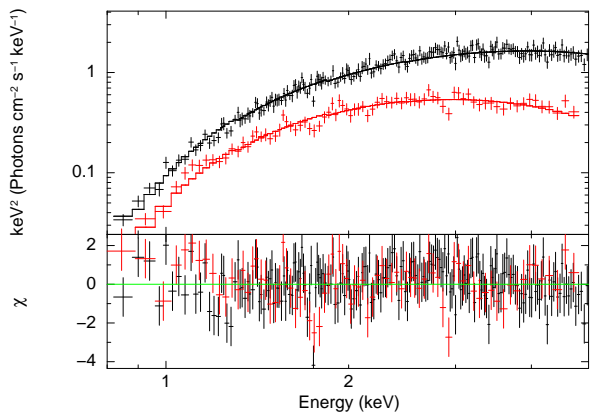}
    \caption{Count rate  resolved spectra of the observation 00031921030. The upper spectrum was extracted during time intervals corresponding to the peaks of the flare-like events observed in this observation. The lower spectrum corresponds to the time intervals of the flares where the source count rate was a minimum. The two spectra were fit together with unabsorbed {\tt MDBB} model (we constrained the N$_{H}$ to be the same  for the two spectra and we let the other parameters to vary independently).}
         \label{spepick}
\end{figure}

The unfolded phase-resolved spectra obtained for the XRT observation n$^{o}$30 (MJD=55640.5) are shown in 
Figure~\ref{spepick}  .
We found evidence  that the flares are due to an oscillation of the inner disc boundary (Table~\ref{tab_peak}): at the peak of the flare 
the {\tt MDBB} temperature (radius) is higher (smaller) with the disc approaching the BH event horizon. The opposite behaviour  is 
observed during the minima of the flare.  This is similar to what has been observed in the case of GRS~1915$+$105~\citep{Neilsen}. 
 The lower X-ray flux of IGR~J17091$-$3624 with respect to GRS~1915$+$105, however, does not allow us to study the ``heartbeat'' in the same details.
  Theoretical studies suggest that this phenomenon is due to the Lightman-Eardley instability, a limit cycle in the inner accretion disc 
dominated by the radiation-pressure~\citep{Lightman, Nayakshin, Szuszkiewicz}.  According to this interpretation, 
the inner part of the disc empties and refills with a timescales of seconds~\citep{Belloni97}. 

\begin{table*}
 \begin{center}
\caption{Spectral parameters of the different phases of three XRT observations: N is the number of the XRT observation as in Table~\ref{parameters}; {\it H}: maxima count rate intervals ($>$ 60 ct/s); {\it L}: minima count rate intervals ($<$ 30 ct/s).}
 \label{tab_peak}
\leavevmode
 \begin{tabular}{lccccccc}
 N & ID & Phase         & T$_{in}$ & NORM & F$_{(0.1-10keV)}$      &$\chi^{2}_{red.}$ & d.o.f.\\
  -&  - & -             & keV      &{\it diskbb}&  ($\times$10$^{-9}$erg~cm$^{-2}$s$^{-1}$)   & -     & -\\
\hline
26& 00031921028 & {\it H}  & 1.4$^{+0.1}_{-0.1}$&69$^{+12 }_{-10}$ & 6 &1.04& 67\\
26& 00031921028 & {\it L}  & 1.1$^{+0.1}_{-0.1}$&81$^{+27 }_{-21}$& 2 &1.17& 24\\
\hline
28& 00031921030 & {\it H}  & 1.49$^{+0.03}_{-0.03}$  &52$^{+4 }_{-4}$&  6  &0.99& 212\\
28& 00031921030 & {\it L}  & 1.10 $^{+0.03}_{-0.03}$ &63$^{+8 }_{-7}$& 2 &1.02& 83\\
\hline
38& 00031921042 & {\it H}  & 1.6$^{+0.1}_{-0.1}$     &37 $^{+5 }_{-4}$& 5  &0.88& 112\\
38& 00031921042 & {\it L}  & 1.00$^{+0.02}_{-0.02}$ &83$^{+8 }_{-7}$& 2 &0.98& 128\\
\hline
51& 00035096002 &{\it H}& 1.4$^{+0.1}_{-0.1}$ & 52$^{+12 }_{-10}$ & 4 & 0.85 &42\\
51& 00035096002 &{\it L} & 1.18$^{+0.03}_{-0.03}$ & 46$^{+5 }_{-4}$&2& 1.00& 126\\
\hline
54& 00035096005 &{\it H} & 1.5$^{+0.1}_{-0.1}$ & 38$^{+8 }_{-7}$ & 4 & 1.3 &64\\
54& 00035096005 &{\it L} & 1.23$^{+0.03}_{-0.03}$ & 40$^{+3 }_{-3}$&2& 1.10& 185\\
 \hline
 \end{tabular}
\end{center}
 \end{table*}

\subsection{Reflection component}
\label{reflection}

   In order to  investigate the presence of Compton-reflection component and the iron line in the spectra of IGR~J17091$-$3624, we used the XRT, JEM-X and IBIS joint spectra 
showed in Figure~\ref{newspe}. In this case the spectral parameters revealed that IGR~J17091$-$3624 is in the soft state (observation n$^{o}$33) when the highest contribution 
from the reflection component is expected~\citep[see e. g.][and reference therein]{Ross}. The model used to fit the data is an absorbed 
{\tt MDBB} plus an exponentially cut-off power-law spectrum reflected by neutral material~\citep[{\it pexrav} in XSPEC;][]{Magdziarz}.

Considering the distance of the source estimated by ~\citet{Pahari} and ~\citet{Rodriguez}, we took into account also the hypothesis that the source could belong 
to the Galactic halo and thus have a different metallicity  with respect to the sources in the Galactic bulge, where normally   LMXBs are concentrated~\citep{Grimm}.
 No significant changes in the spectral fits were observed by leaving the metallicity of the reflecting medium free to vary. We thus assumed 
two different values of the metallicity, i.e. the solar one (the source belongs to  Galactic bulge, Z/Z$_{\odot}$ = 1) and Z/Z$_{\odot}$ = 0.13 as reported 
by~\citet{Frontera} for XTE J1118+480 which is a BH binary that lies at very high Galactic Latitudes.
 In both cases the estimated upper limit on the reflection component was of  R=0.1,  and the F-test 
 probability indicated that there is not a clear evidence of a significant improvement in the $\chi^{2}$ by adding this component 
 (the F-test probability in the two cases was of 7\% and 2\%, corresponding to 
a detection significance of $<$2.0 $\sigma$ and $<$2.5$\sigma$, respectively).

{We also estimated an upper limit on the normalization of the iron line fixing the centroid of the line at 6.7~keV. 
We assumed a broad line with $\sigma$=0.7 keV as in the case of GRS~1915+105~\citep[see e. g.][and references therein]{Martocchia}. 
The obtained upper limit on the equivalent width is EQ$<$ 0.9~keV.}

\section{Discussion}
\label{disc}
All the outbursts of IGR~J17091$-$3624 observed before 2011 were fainter and poorly observed with respect to the last one. However the source, in the limit of the instruments capability, displayed the typical spectral and temporal 
evolution~\citep{Cap,Cap2} expected from a canonical BHC \citep[for details on the transient BHC outburst evolution see e.g.,][]{Fender}. 
The  ``heartbeat'' phenomenon appeared only during the last 2011  outburst.  Indeed, using all the available archival XRT observations in the direction of IGR~J17091$-$3624,   we verified that no ``heartbeat'' was visible during the previous outbursts of the source.

 We summarize here the initial evolution phases of the outburst occurred on 2011.
The source underwent to a transition from the LHS to the HSS moving from the bottom right corner of the HID to the 
top left corner (Figure~\ref{HID}, observations n$^{o}$1$\div$15):
\begin{itemize}
\item during this transition,   the source reached the intermediate states and the radio flare reported by~\citet{Rodriguez} 
should be the signature of the transition from the HIMS to the SIMS~\citep{Fender};
\item the {\it rms} amplitude   starting from values of about $\sim$30\% in the LHS, decreased significantly
reaching values that  span from 6\% to 2\%  (see Figure~\ref{RMS} and column n$^{o}$6 in Table~\ref{parameters});
\item  the spectrum became softer   with the presence of a prominent disc  blackbody component  
(starting from observation n$^{o}$15)   with the high energy cutoff   no longer detectable up to 200 keV.
\end{itemize}
The source remained in the HSS for about 10 days (from MJD=55623.5 till MJD=55633.3). 
Starting from MJD=55635 the source followed no more the standard evolution of a transient BHC in outburst:
the  properties of the X-ray spectra in each observation showed no significant variability, while the source displayed a sudden atypical 
timing variability in the form of flare-like events occurring at a 33\,s period (``heartbeat''). 
{ The X-ray emission at the peak of these flares is typically harder than the average source emission (see the third panel of Figure~\ref{hbeats}).

Starting from MJD=55692 we measured a progressive decrease of the {\tt MDBB} inner temperature with a corresponding hardening of the source emission.  
At this time the flare-like events were no longer visible in the light curve. The hardening continued uninterrupted for about two days, 
then the inner temperature of the disc started to increase again, leading to a  clear increase in the soft X-ray flux  and a decrease of the hard X-ray emission.  
At this epoch, the ``heartbeat'' became again visible.}

The last part of the data analysed presented a   short period oscillations (between 3\,s and 30\,s) and also 
a particularly hot inner disc temperature with a very small {\tt MDBB} normalization constant that corresponds to a small apparent inner radius.
Between MJD=55740 and MJD=55760, the 4-10 keV XRT flux increased significantly (a factor of 60\%) . 
The peak in the 4-10 keV flux (see Figure~\ref{xrtlctot}) corresponds to a peak in the inner disc temperature (T$_{in}\sim$1.7 keV on MJD=55759.3).

 The period of the ``heartbeat'' changed with time   (Figure~\ref{periods}) and it  seems to have a decreasing trend. 

\subsection{Comparison with the BH binary GRS~1915$+$105}
\label{1915}
 As reported by~\citet{Altamirano} and by~\citet{Pahari} the behaviour of the source resembles what observed from  GRS~1915$+$105 in the various flaring states.
Thus the principal common characteristic between these two sources is just the presence of pseudo periodic flare-like events in the light curve  i. e. the so called  ``heartbeat''. 
The  HR (bottom panel of Figure~\ref{hbeats}) of IGR~J17091$-$3624 is similar to the GRS~1915$+$105 one, in the sense that in both sources  the modulation of the light curve is projected also in the HR~\citep{Neilsen}.  However, in the GRS~1915$+$105 case the hardness variation seems more pronounced  \citep[see for example][]{Naik}.
 Our phase resolved energy spectra of the XRT data revealed that the hardening of the source X-ray emission at the peak and at the lower part of each flare is similar to what measured in the case of GRS~1915$+$105~\citep[see e. g.][]{Mineo, Belloni97} and thus it is probably due to the same physical phenomenon~\citep{Lightman}. 

The period of the ``heartbeat'' seems also to vary with time in the same range of values for the two sources, even though 
in GRS~1915$+$105 the period amplitude gets larger for long time scales. 
  This does not seem to be the case for IGR~J17091$-$3624. Indeed, as showed in  Figure~\ref{periods}, 
 the period variation with time seems to decrease and, moreover, in the third group of observations 
 (from  MJD=55750 until the end) it reaches values of the order of few seconds ($\sim$3$\div$5\,s). 
 These values were not observed in GRS~1915$+$105~\citep{Neilsen}. 

 Similarly to GRS~1915$+$105, we measured also for IGR~J17091$-$3624   particularly hot inner disc  temperatures \citep[in the case of GRS~1915$+$105 the temperature can reach even higher values; see e.g.,][]{Belloni97,Muno, Fender2}~\footnotemark\footnotetext[6]{ The inner disc radius values reported for GRS~1915+105 by~\citet{Muno} and related to inner temperatures greater than 1.6 keV, are too small to be associated with the ISCO for any reasonable black hole mass. Even if the hard part of the spectrum, modeled using a power law, could underestimate the inner disc radius~\citep{Done}, it is not possible to exclude that, in these cases, the accretion geometry could be different from the one predicted by {\tt MDBB}. However this should not be the case for IGR J17091-3624. In fact, the spectral parameters reported in our analysis are not as extreme as the ones reported by \citet{Muno}  for GRS~1915+105.}. 
   This property, together with a small inner radius  of the disc  blackbody spectrum in X-ray binaries, 
has been directly associated with high values of the BH spin~\citep{Zhang,Devis}. 

 Besides all these similarities between GRS~1915$+$105 and IGR~J17091$-$3624, a particularly striking  difference is the  X-ray flux intensity during the outbursts.
 This fact cannot be easily explained because, unlike GRS~1915$+$105, for IGR~J17091$-$3624 we do not have an estimation
 of the distance, the inclination angle, BH mass and spin, and the properties of the companion star. Some results on optical  
 and NIR counter-part of IGR~J17091$-$3624 have been reported by~\citet{Atel3150}.

\citet{Chaty},  on the basis of optical and NIR photometric and spectroscopic  studies of two possible counter-parts of the source, 
 suggested that the source should belong to the Galactic bulge. However, 
~\citet{Rodriguez} recently estimated  a lower limit of the source distance from its hard to soft transition luminosity concluding that, 
if the transition occurred at luminosity that spans from 4\% to 10\% of the Eddington luminosity (assuming a BH mass 
of 10 M$_{\odot}$), IGR~J17091$-$3624 is farther from the Galactic bulge, at a distance that spans from  about 11\,kpc up to 17\,kpc. 
Moreover \citet{Pahari}, using a different method, based on QPO, estimated an even
larger distance of 20\,kpc and a mass range that spans from 8M$_{\odot}$ to 11.4M$_{\odot}$.

Assuming  a distance range  of 11-17 kpc,  the  bolometric luminosities of IGR~J17091$-$3624  estimated  from the observation displaying ``heartbeat'' with the highest flux would be  (3-7)$\times$ 10$^{37}$ erg~s$^{-1}$ which translates in  to L$\sim$(3-6)\% L$_{Edd}$.
However, considering the distance and the BH mass range supposed by~\citet{Pahari} these luminosities result in 1\% and 8\% L$_{Edd}$. 

 Since the flare-like events should be at Eddington limit regime~\citep[see e.g.][and reference therein]{Nayakshin,Neilsen}, if we consider the values reported above, we conclude that the faintness of IGR~J17091$-$3624 should not  be only due to the source distance. For this reason ~\citet{Altamirano} supposed that the distance of the source could be even larger  than 20\,kpc, otherwise the BH mass should be extremely small (less than 3$M_{\odot}$). 

 Other  peculiar differences of IGR~J17091$-$3624 with respect to 
 GRS~1915$+$105 are the lack of detection of the Compton-reflection component and
  the extremely low apparent inner disc radius  (see Section~\ref{heartbeat_p}). 

 Taking these results as a whole, we speculate that IGR~J17091$-$3624  could be a  highly inclined system and we suggest that the lower luminosity of IGR~J17091$-$3624 could be   also ascribed to the spectral deformation effects due to the high inclination angle  as reported by ~\citet{Cunningham}. Indeed, when a Kerr BH is seen at a high inclination angle (cos{\tt i} $<$0.25, {\tt i}$\sim$75 degrees), the source appears significantly fainter (up to a factor that depends on  the BH spin and mass but can reach about an order of magnitude less) with respect to system observed  face-on. 

  At odds  with this  hypothesis is the lack of  detection of eclipses. Although we do not have any information about the system, such as the orbital period or the companion star mass, we can speculate that the lack of eclipses  could be related to a small ratio between the companion star and BH mass. 
Using the Eggleton approximation~\citep{Eggleton}, we calculated the relation between the mass ratio, $q$ (M$_{star}$/M$_{BH}$) and the Roche lobe radius.
 Then, the minimum inclination angle, $i$, for which the Roche lobe does not cover the central engine along the observer line of sight, is extracted from simple geometrical considerations giving:
\begin{equation}
\label{eq1}
R_{L}/a < cos{\tt i}
\end{equation}
where R$_{L}$ is the Roche Lobe radius calculated with the Eggleton approximation;
$a$ is the distance between the BH and the companion star;
 $i$ is the inclination angle of the system.
Plotting R$_{L}/a$ versus $q$~\citep{Eggleton} and considering the equation~\ref{eq1}, we found that  for $q<0.2$, the Roche lobe does not cover the central engine for inclination angles smaller than 75 degrees~\citep{Cunningham}.

Moreover,  the lack of  information on the orbital period of the system hampers the search for   the eventual presence of partial eclipses via the usual light curve folding techniques  that increase the signal to noise ratio.

\section{Conclusions}
\label{conclusions}
 The outcome of the observational campaign presented here suggests that IGR~J17091$-$3624 can be no longer considered  as typical transient 
Black Hole~\citep{Fender}. After the transition from hard to soft state  in 2011~\citep{Rodriguez}, the source  did not follow the standard  
 q-track in the HID diagram \citep[see e. g.][and reference therein]{Homan,Homan2005} and, since March 2011, it remained trapped in an oscillatory state, similarly to what 
observed during the flaring states of GRS~1915$+$105~\citep{Altamirano}.

As mentioned above (see section~\ref{1915}), the pseudo periodic bursts in the light curve  of GRS~1915$+$105 reach the Eddington luminosity and are believed to be related to disc oscillations. The physics that drives these inner disc oscillations is connected with both  the local Eddington limit and the radiation pressure instability.
If the ``heartbeat'' oscillations  seen from IGR~J17091$-$3624 are interpreted as being due to the same mechanism as in GRS~1915$+$105 ~\citep[as also supposed by][]{Altamirano}, then the apparent "faintness" of IGR~J17091$-$3624 remains   unexplained unless to suppose a huge distance or an extremely low BH mass~\citep{Altamirano}.

 In Section~\ref{1915} we noted that that a reduction of the apparent luminosity up to an order of magnitude can 
also be achieved if the system is seen nearly edge on~\citep[for inclination angles $<$75 degrees,][]{Eggleton}. 
According to this idea and considering also the L/L$_{Edd}$ ratio calculated for the different distance values, we can speculate that the source,  probably, not only lies far from the Galactic bulge, in agreement with ~\citet{Rodriguez}, but it is observed at a high inclination angle as well. 
  As also discussed in Section~\ref{1915}, this finding is not in contrast with the lack of eclipses in the source light curve. In fact,  if the companion star is  small, the eclipses  can be undetected even for a high inclination angle, as for example in the case of the BHC XTE J1118+480~\citep[see e.g.][]{Wagner, McClintock}.

 We note that at present we cannot exclude that the faintness of IGR~J17091$-$3624 is only due to a very 
 large distance ($>$20 kpc) or to the extremely low BH mass ($<$ 3M$_{\odot}$), as suggested by \citet{Altamirano}.
The large distance, unusual for  low mass X-ray binaries  generally concentrated in the Galactic  bulge~\citep{Grimm}, 
could agree with the hypothesis reported by~\citet{Jonker} that the distances of the LMXB could be affected by 
a systematic error due to misclassification of the companion star.

 However recent results, reported by~\citet{King}, based on a  {\it Chandra} observation campaign, support the hypothesis that IGR~J17091$-$3624 is observed at high inclination angle.
 Future refined estimation of the distance and the BH mass of IGR~J17091$-$3624 might help understand if 
 GRS~1915$+$105 and IGR~J17091$-$3624 are very similar objects simply observed at very different distances or inclination angles.
We point out that the high inclination of the system 
is a possible scenario to explain the low luminosity of the source without invoking very large distances or extremely low 
BH masses that may challenge the Rhoades \& Ruffini limit~\citep{Ruffini}.

Finally we suggest that, as  in the case of GRS~1915$+$105, also IGR~J17091$-$3624 might show a "quasi-persistent" outburst of the order of years.
Thus the {\it INTEGRAL} and {\it Swift} observation  campaign of the 2011 outburst probably caught the evolution of a transient BH in a persistent GRS~1915$+$105-like phase.

\section*{Acknowledgments}
FC, MDS, AP and GDC acknowledge financial support from the agreement ASI-INAF I/009/10/0. 
FC thanks Giorgio Matt and Piergiorgio Casella for useful scientific discussions.
MDS and GDC acknowledge contribution by the grant PRIN-INAF 2009.
We would like to thank N. Gehrels and the {\it Swift} Team for making {\it Swift} observations possible. 
A special thanks goes to E. Kuulkers and the {\it INTEGRAL} Galactic bulge monitoring program.
{\it INTEGRAL} is an ESA project with instruments and science data centre funded by ESA member states especially the PI countries: Denmark, France, Germany, Italy, Switzerland, Spain, Czech Republic and Poland, and with the participation of Russia and the USA.

\end{document}

%% file: tabella_referee.tex
1&  00031921002&  55598.3&  1940&  1016&  0.29$\pm$0.08&   -&  -&  1.4$_{-0.1}^{+0.1}$&  109$_{-19}^{+28}$&  6.1&  1.0(139)\\
2&  00031921003&  55599.2&  2170&  1016&  0.28$\pm$0.03&   -&  -&  1.5$_{-0.1}^{+0.1}$&  120$_{-21}^{+30}$&  6.5&  1.0(193)\\
3&  00031921004&  55600.7&  2175&  1016&  0.25$\pm$0.07&   -&  -&  1.5$_{-0.1}^{+0.1}$&  104$_{-14}^{+19}$&  6.8&  1.0(164)\\
4&  00031921005&  55601.1&  1454&  -&  0.25$\pm$0.06&   -&  -&  1.5$_{-0.1}^{+0.1}$&  -&  6.1&  0.8/175)\\
5&  00031921006&  55602.1&  2191&  1017&  0.21$\pm$0.03&   -&  -&  1.6$_{-0.1}^{+0.1}$&  125$_{-21}^{+31}$&  9.3&  1.0(174)\\
6&  00031921007&  55603.2&  2066&  1017&  0.24$\pm$0.06&   -&  -&  1.5$_{-0.1}^{+0.1}$&  86$_{-9}^{+11}$&  8.6&  1.0(210)\\
7&  00031921008&  55604.2&  2189&  1017&  0.27$\pm$0.03&   -&  -&  1.5$_{-0.1}^{+0.1}$&  93$_{-9}^{+11}$&  9.4&  1.3(264)\\
8&  00031921009&  55605.2&  2163&  1018&  0.24$\pm$0.07&   -&  -&  1.5$_{-0.1}^{+0.1}$&  86$_{-21}^{+35}$&  10.0&  1.1(200)\\
9&  00031921010&  55606.2&  1884&  -&  0.22$\pm$0.03&   -&  -&  1.6$_{-0.1}^{+0.1}$&  -&  8.6&  1.1(139)\\
10&  00031921011&  55607.2&  2108&  -&  0.24$\pm$0.05&   -&  -&  1.6$_{-0.1}^{+0.1}$&  -&  8.9&  1.1(217)\\
11&  00031921012&  55608.3&  2057&  -&  0.22$\pm$0.02&   -&  -&  1.61$_{-0.04}^{+0.04}$&  -&  9.1&  1.1(249)\\
12&  00031921013&  55610.2&  2010&  -&  0.21$\pm$0.03&   -&  -&  1.6$_{-0.1}^{+0.1}$&  -&  10.4&  1.0(225)\\
13&  00031921014&  55612.3&  2095&  1020&  0.15$\pm$0.05&   -&  -&  1.69$_{-0.04}^{+0.04}$&  75$_{-7}^{+9}$&  14.7&  1.0(311)\\
14&  00031921015&  55614.2&  2195&  1020&  0.08$\pm$0.04&   0.3$_{-0.1}^{+0.1}$&  $<$1$\times$10$^{5}$&  2.0$_{-0.1}^{+0.1}$&  134$_{-29}^{+43}$&  20.5&  1.2(366)\\
15&  00031921016&  55616.3&  1074&  -&  0.05$\pm$0.02&   1.1$_{-0.1}^{+0.1}$&  53$_{-24}^{+38}$&  2.1$_{-0.5}^{+0.3}$&  -&  20.6&  1.1(322)\\
16&  00031921017&  55620.8&  2568&  -&  0.06$\pm$0.02&   1.0$_{-0.1}^{+0.1}$&  54$_{-25}^{+35}$&  2.1$_{-0.2}^{+0.2}$&  -&  17.8&  1.0(428)\\
17&  00031921018&  55622.5&  2321&  -&  0.09$\pm$0.03&   1.0$_{-0.1}^{+0.2}$&  46$_{-37}^{+63}$&  2.1$_{-0.4}^{+0.3}$&  -&  19.5&  1.2(268)\\
18&  00031921019&  55623.5&  656&  -&  0.05$\pm$0.02&   1.1$_{-0.1}^{+0.2}$&  54$_{-27}^{+41}$&  2.1$_{-0.4}^{+0.3}$&  -&  18.2&  1.1(316)\\
19&  00031921020&  55624.4&  2463&  -&  0.05$\pm$0.02&   1.1$_{-0.1}^{+0.1}$&  60$_{-18}^{+18}$&  1.7$_{-0.7}^{+0.3}$&  -&  18.9&  1.1(443)\\
20&  00031921021&  55627.6&  2072&  1025&  0.04$\pm$0.02&   1.3$_{-0.1}^{+0.1}$&  49$_{-14}^{+24}$&  2.4$_{-0.1}^{+0.1}$&  -&  35.1&  1.0(433)\\
21&  00031921022&  55628.1&  1706&  1025&  0.05$\pm$0.02&   1.29$_{-0.04}^{+0.04}$&  98$_{-30}^{+56}$&  2.4$_{-0.1}^{+0.2}$&  -&  47.6&  1.1(489)\\
22&  00031921023&  55630.5&  1408&  -&  0.05$\pm$0.02&   1.03$_{-0.04}^{+0.05}$&  90$_{-19}^{+13}$&  1.3$_{-0.1}^{+0.5}$&  -&  17.1&  1.2(401)\\
23&  00031921024&  55632.3&  2189&  1027&  0.07$\pm$0.02&   1.29$_{-0.04}^{+0.04}$&  71$_{-21}^{+33}$&  2.6$_{-0.1}^{+0.1}$&  -&  40.3&  1.0(411)\\
24&  00031921025&  55633.3&  2016&  -&  0.08$\pm$0.06&   1.31$_{-0.02}^{+0.02}$&  51$_{-4}^{+4}$&  -&  -&  18.2&  1.2(357)\\
25&  00031921026&  55635.6&  1473&  1028&  0.07$\pm$0.03&   1.2$_{-0.1}^{+0.1}$&  35$_{-8}^{+12}$&  2.1$_{-0.1}^{+0.1}$&  -&  15.2&  1.2(434)\\
26&  00031921028&  55639.8&  2225&  -&  0.27$\pm$0.02&   1.28$_{-0.03}^{+0.03}$&  63$_{-26}^{+20}$&  -&  -&  20.7&  1.0(274)\\
27&  00031921029&  55638.6&  2155&  -&  0.10$\pm$0.02&   1.29$_{-0.02}^{+0.02}$&  54$_{-3}^{+3}$&  -&  -&  18.7&  1.0(417)\\
28&  00031921030&  55640.5&  2348&  -&  0.28$\pm$0.02&   1.29$_{-0.02}^{+0.02}$&  57$_{-3}^{+3}$&  -&  -&  19.2&  1.3(442)\\
29&  00031921031&  55642.1&  2137&  1030&  0.26$\pm$0.01&   1.32$_{-0.03}^{+0.03}$&  47$_{-4}^{+4}$&  -&  -&  25.6&  1.0(308)\\
30&  00031921033&  55643.4&  630&  1030&  0.11$\pm$0.06&   1.2$_{-0.1}^{+0.1}$&  64$_{-31}^{+138}$&  2.1$_{-0.2}^{+0.3}$&  -&  23.6&  1.1(314)\\
31&  00031921034&  55646.9&  2166&  -&  0.12$\pm$0.02&   1.29$_{-0.02}^{+0.02}$&  45$_{-3}^{+3}$&  -&  -&  15.0&  1.2(382)\\
32&  00031921036&  55650.2&  326&  -&  0.192$\pm$0.003&   1.28$_{-0.04}^{+0.04}$&  51$_{-6}^{+6}$&  -&  -&  16.9&  0.9(196)\\
33&  00031921037&  55651.5&  2067&  1033&  0.22$\pm$0.02&   1.2$_{-0.1}^{+0.1}$&  55$_{-25}^{+91}$&  2.1$_{-0.3}^{+0.2}$&  -&  17.4&  1.0(288)\\
34&  00031921038&  55654.7&  1173&  1034&  0.31$\pm$0.01&   1.27$_{-0.10}^{+0.04}$&  74$_{-41}^{+163}$&  2.4$_{-0.3}^{+0.2}$&  -&  29.6&  1.2(424)\\
35&  00031921039&  55655.7&  1022&  -&  0.31$\pm$0.02&   1.31$_{-0.02}^{+0.02}$&  52$_{-4}^{+4}$&  -&  -&  18.8&  1.1(397)\\
36&  00031921040&  55657.5&  2111&  1035&  0.31$\pm$0.01&   1.22$_{-0.10}^{+0.09}$&  57$_{-30}^{+140}$&  2.6$_{-0.4}^{+0.4}$&  -&  23.4&  0.8(216)\\
37&  00031921041&  55661.9&  974&  -&  0.32$\pm$0.02&   1.31$_{-0.02}^{+0.02}$&  45$_{-3}^{+3}$&  -&  -&  16.6&  1.1(382)\\
38&  00031921042&  55667.3&  1069&  -&  0.34$\pm$0.04&   1.29$_{-0.02}^{+0.02}$&  48$_{-3}^{+3}$&  -&  -&  16.7&  1.2(387)\\
39&  00031921043&  55679.8&  896&  -&  0.33$\pm$0.02&   1.29$_{-0.02}^{+0.02}$&  48$_{-3}^{+4}$&  -&  -&  16.1&  1.2(359)\\
40&  00031921044&  55681.2&  1035&  -&  0.26$\pm$0.04&   1.29$_{-0.02}^{+0.02}$&  44$_{-3}^{+3}$&  -&  -&  15.0&  1.1(375)\\
41&  00031921045&  55683.7&  1509&  -&  0.15$\pm$0.03&   1.25$_{-0.02}^{+0.02}$&  52$_{-3}^{+3}$&  -&  -&  15.1&  1.2(397)\\
42&  00031921046&  55685.3&  1246&  -&  0.05$\pm$0.01&   1.26$_{-0.03}^{+0.03}$&  49$_{-5}^{+5}$&  -&  -&  13.2&  1.1(231)\\
43&  00031921049&  55691.6&  2092&  -&  0.08$\pm$0.06&   1.1$_{-0.1}^{+0.2}$&  37$_{-16}^{+31}$&  2$_{-2}^{+1}$&  -&  10.9&  0.9(198)\\
44&  00031921050&  55693.1&  2318&  -&  0.08$\pm$0.04&   1.0$_{-0.1}^{+0.1}$&  62$_{-15}^{+9}$&  1$_{-2}^{+1}$&  -&  9.7&  1.0(311)\\
45&  00031921051&  55695.0&  1139&  -&  0.08$\pm$0.03&   1.2$_{-0.2}^{+0.2}$&  15$_{-9}^{+19}$&  2.4$_{-0.3}^{+0.3}$&  -&  9.1&  1.0(303)\\
46&  00031921052&  55697.8&  1160&  -&  0.07$\pm$0.04&   1.1$_{-0.1}^{+0.2}$&  26$_{-12}^{+23}$&  2.4$_{-0.4}^{+0.3}$&  -&  8.7&  1.2(298)\\
47&  00031921053&  55701.8&  1188&  -&  0.10$\pm$0.02&   1.1$_{-0.1}^{+0.2}$&  20$_{-10}^{+19}$&  2.4$_{-0.4}^{+0.3}$&  -&  8.6&  1.1(300)\\
48&  00031921054&  55703.7&  2291&  -&  0.08$\pm$0.02&   1.0$_{-0.1}^{+0.2}$&  30$_{-19}^{+23}$&  -&  -&  9.6&  1.0(299)\\
49&  00031921055&  55705.6&  2374&  -&  0.15$\pm$0.03&   1.23$_{-0.02}^{+0.02}$&  49$_{-3}^{+4}$&  -&  -&  13.6&  1.1(375)\\
50&  00031921056&  55707.4&  1789&  -&  0.26$\pm$0.02&   1.24$_{-0.02}^{+0.02}$&  50$_{-3}^{+3}$&  -&  -&  14.4&  1.1(386)\\
51&  00035096002&  55715.7&  894&  -&  0.18$\pm$0.04&   1.27$_{-0.02}^{+0.02}$&  47$_{-4}^{+4}$&  -&  -&  14.7&  1.3(345)\\
52&  00035096003&  55717.7&  1006&  -&  0.12$\pm$0.02&   1.20$_{-0.02}^{+0.02}$&  46$_{-3}^{+4}$&  -&  -&  11.3&  1.3(334)\\
53&  00035096004&  55719.5&  907&  -&  0.13$\pm$0.05&   1.23$_{-0.02}^{+0.02}$&  45$_{-3}^{+4}$&  -&  -&  12.3&  1.1(323)\\
54&  00035096005&  55721.3&  1070&  -&  0.17$\pm$0.02&   1.24$_{-0.02}^{+0.02}$&  49$_{-37}^{+4}$&  -&  -&  13.9&  1.1(369)\\
55&  00035096009&  55725.6&  1151&  -&  0.06$\pm$0.03&   1.20$_{-0.02}^{+0.02}$&  46$_{-3}^{+4}$&  -&  -&  9.0&  1.3(334)\\
56&  00035096010&  55729.4&  1006&  -&  0.07$\pm$0.04&   1.3$_{-0.3}^{+0.3}$&  8$_{-3}^{+13}$&  2.5$_{-0.4}^{+0.9}$&  -&  7.8&  1.0(247)\\
57&  00035096012&  55731.0&  437&  -&  0.07$\pm$0.05&   1.09$_{-0.02}^{+0.04}$&  18$_{-12}^{+42}$&  2$_{-1}^{+1}$&  -&  8.1&  0.8(132)\\
58&  00035096014&  55733.2&  928&  -&  0.11$\pm$0.05&   1.0$_{-0.1}^{+0.1}$&  43$_{-19}^{+26}$&  2$_{-5}^{+1}$&  -&  7.8&  1.0(238)\\
59&  00035096015&  55735.3&  1102&  -&  0.10$\pm$0.03&   1.1$_{-0.1}^{+0.1}$&  49$_{-13}^{+30}$&  3$_{-1}^{+1}$&  -&  12.5&  1.1(283)\\
60&  00035096016&  55737.5&  140&  -&  0.16$\pm$0.04&   1.2$_{-0.1}^{+0.1}$&  51$_{-14}^{+19}$&  -&  -&  11.8&  1.0(46)\\
61&  00035096017&  55739.4&  992&  -&  0.32$\pm$0.05&   1.2$_{-0.03}^{+0.03}$&  33$_{-3}^{+4}$&  -&  -&  8.1&  0.9(224)\\
62&  00035096018&  55741.6&  1123&  -&  0.39$\pm$0.03&   1.47$_{-0.03}^{+0.03}$&  25$_{-2}^{+2}$&  -&  -&  15.2&  1.3(375)\\
63&  00035096019&  55744.0&  984&  -&  0.37$\pm$0.02&   1.58$_{-0.04}^{+0.04}$&  19$_{-2}^{+2}$&  -&  -&  17.2&  1.1(350)\\
64&  00035096020&  55759.3&  850&  -&  0.36$\pm$0.05&   1.67$_{-0.05}^{+0.06}$&  18$_{-2}^{+2}$&  -&  -&  19.6&  1.1(210)\\
65&  00035096021&  55761.5&  727&  -&  0.42$\pm$0.03&   1.50$_{-0.03}^{+0.04}$&  24$_{-2}^{+2}$&  -&  -&  16.1&  1.2(311)\\
66&  00035096022&  55765.3&  956&  -&  0.41$\pm$0.04&   1.35$_{-0.03}^{+0.03}$&  36$_{-3}^{+3}$&  -&  -&  15.0&  1.0(363)\\
67&  00035096023&  55767.3&  940&  -&  0.38$\pm$0.04&   1.32$_{-0.02}^{+0.02}$&  37$_{-3}^{+3}$&  -&  -&  14.0&  1.1(346)\\
68&  00035096027&  55775.3&  863&  -&  0.39$\pm$0.03&   1.34$_{-0.02}^{+0.03}$&  38$_{-3}^{+3}$&  -&  -&  15.3&  1.1(342)\\
69&  00035096028&  55777.9&  354&  -&  0.22$\pm$0.01&   1.27$_{-0.04}^{+0.04}$&  47$_{-6}^{+7}$&  -&  -&  14.9&  1.2(165)\\
70&  00035096029&  55779.4&  511&  -&  0.23$\pm$0.06&   1.24$_{-0.03}^{+0.04}$&  50$_{-5}^{+6}$&  -&  -&  14.3&  1.0(237)\\
71&  00035096030&  55783.8&  547&  -&  0.34$\pm$0.03&   1.30$_{-0.04}^{+0.04}$&  82$_{-10}^{+11}$&  -&  -&  29.0&  1.1(198)\\
 -&            -&  55785.0&    -&  1078&           -&   1.3$_{-0.1}^{+0.1}$&  58$_{-20}^{+32}$&  2.3$_{-0.2}^{+0.2}$&  -&  24.7&  1.1(26)\\
72&  00035096032&  55787.7&  974&  -&  0.37$\pm$0.04&   1.28$_{-0.02}^{+0.02}$&  58$_{-5}^{+5}$&     -&  -&  19.2&  1.1(352)\\    